\documentstyle[prd,aps,epsf,eqsecnum]{revtex}

\begin{document}
\draft
\title{Energy-Momentum Tensor and 
Particle Creation in the de Sitter Universe}

\author{Carmen Molina-Par{\'{\i}}s
\thanks{electronic mail: carmen@t6-serv.lanl.gov}}
\address{Los Alamos National Laboratory, Theoretical Division, 
 Los Alamos, NM, 87544}
\date{\today}
\maketitle

\begin{abstract}{
Particle creation in a conformally flat spacetime ({\rm e.g.,} FRW
universe) requires a non-conformal field.  The choice of state is
crucial, as one may misunderstand the physics of particle creation by
choosing a too restrictive vacuum for the quantum field. We exhibit a
vacuum state in which the expectation values of the energy and
pressure allow an intuitive physical interpretation. We apply
this general result to the de Sitter universe.  }
\end{abstract}
\section{Introduction}
\label{sec:intro}

We first consider a charged scalar field $\Phi$ in Minskowski spacetime.
Suppose that
the electric field is ${\bf E} = E {\bf z}$, and the vector
potential is ${\bf A}(t) = - E t {\bf z}$. The wave equation for a charged 
field in a Minkowski universe is given by:
\begin{eqnarray}
[{(\partial_\mu - i e A_\mu)}^2 + m^2] \Phi(t,{\bf x})
 = 0
\>.
\end{eqnarray}
We can find solutions of the form (Fourier mode decomposition)
\begin{eqnarray}
\Phi(t,{\bf x}) = {1 \over {V^{1/2}}} \sum_{\bf k}
a_{\bf k} f_{\bf k} (t) e^{i {\bf k} \cdot {\bf x}}
+
b_{- {\bf k}}^{\dag} f_{-{\bf k}}^{*} (t) e^{-i {\bf k} \cdot {\bf x}}
\>,
\end{eqnarray}
with mode functions that satisfy the following harmonic
differential equation:
\begin{eqnarray}
{\ddot f}_{\bf k}(t) + \omega^2_{\bf k}(t) f_{\bf k}(t) = 0
\; \; \; {\rm where} \; \; \; 
\omega^2_{\bf k}(t)={(k_z+eEt)}^2+ k^2_{\perp} + m^2
\>.
\end{eqnarray}
These mode functions are invariant under time reversal, 
$(t \rightarrow -t, k_z \rightarrow -k_z)$, but the expectation value of
the electric current is odd under this exchange.
This is easy to see; suppose we consider
solutions such that $|f_{\bf k}(t)|=|f_{-{\bf k}}(-t)|$. The expectation
value of the current is
\begin{eqnarray}
\langle j_z  \rangle = {{2e} \over V} \sum_{\bf k}
(k_z + eEt){ |f_{\bf k}(t)|}^2
\> .
\end{eqnarray}
Therefore in the vacuum state defined by
 $a_{\bf k}| 0 \rangle =0=b_{\bf k}| 0 \rangle$, we have
$\langle j_z  \rangle =0 $. On the other hand, we know that 
there are solutions
with adiabatic asymptotic behaviour, such that
\begin{eqnarray}
\lim_{t \rightarrow \pm \infty} f_{{\bf k} (\pm)}(t)
= {\tilde f}_{\bf k}(t) \; \; \; 
{\rm  with} \; \; \; 
{\tilde f}_{\bf k}(t)= { {{[2 \omega_{\bf k}(t)]}^{-1/2}}}
\exp [-i \int^t dt' \omega_{\bf k}(t')]
\> .
\end{eqnarray}
These two families of solutions $\{ f_{{\bf k} (-)}\}$ and 
$\{ f_{{\bf k} (+)}\}$ are related by a Bogoliubov transformation and they
represent two different vacua. If our initial vacuum state is
$|0_{(-)}\rangle$ (adiabatic vacuum at early times), it is easy to
show that in the remote future, when the natural choice for a set of
adiabatic observers is $\{ f_{{\bf k} (+)}\}$, these {\it inertial
observers} would detect particle production given by 
\begin{eqnarray}
N_{{\bf k} (+)}|0_{(-)}
\rangle = [a_{{\bf k} (+)}^{\dag}a_{{\bf k} (+)} + 
b_{-{\bf k} (+)}^{\dag}b_{-{\bf k} (+)}]|0_{(-)} \rangle
\neq 0
\> . 
\end{eqnarray}
These observers will measure a nonvanishing  $\langle  j_z \rangle$.
This is the Schwinger effect which would be completely missed in the
$T-$invariant vacuum state~\cite{Yuval}.

\section{Scalar field in de Sitter spacetime}
\label{sec:desitter}
We next consider a scalar field $\Phi$ in a de Sitter gravitational
background.  We use a coordinate system \cite{Birrel} in which the
spatial sections have curvature $\kappa = +1$, and the scale factor is
$a(t)= Z^{-1}\cosh (Zt)$ with   $u = Zt$.
The wave equation is 
\begin{eqnarray}
[- \Box + m^2 + \xi R] ~  \Phi(t,{\bf x}) = 0
\> .
\end{eqnarray}
In the de Sitter universe
 this equation is separable and the field $\Phi$ can
be written in terms of creation and annihilation operators.
The equation of motion for the mode functions
is
\begin{eqnarray}
\ddot f_k(t) + \Omega^2_k(t) f_k(t) = 0
\> , 
\end{eqnarray}
with
\begin{eqnarray}
\Omega^2_k(t)= Z^2
\left[ \gamma^2 + \left( k+{1 \over 2}
\right) \left( k+{3 \over 2} \right) {\rm sech}^2 (Zt)
\right]
 \; \; \;  {\rm and} \; \; 
 \; \gamma^2 = \frac{m^2}{Z^2} + 12 \left( \xi - {1 \over 6}
\right) - 
{1 \over 4}
\> .
\end{eqnarray}
In the remote past and future
$\Omega_k$ tends to a constant value, and we can find exact mode
functions such
that at early and late times they tend to the adiabatic ones.
These two families of solutions $\{ f_{k (-)}\}$ and $\{ f_{k (+)}\}$
are related by a Bogoliubov transformation
\begin{eqnarray}
f_{k(-)}(t) = \hat
\alpha_k f_{k(+)}(t) + \hat \beta_k f_{k(+)}^*(t)
\> ,
\end{eqnarray}
 where ${|\hat
\beta_k|}^2 = {\rm cosech}^2 (\pi \gamma) \neq 0$. We note that
${|\hat \beta_k|}^2$ is time independent, as well as
$k-$independent. This is the analog of the Schwinger effect mentioned
in the previous section~\cite{Mottola,Gutzwiller}.
Notice that in the de Sitter invariant vacuum there would
no particle creation. This state is the analog of the $T$-invariant
vacuum of the previous section.

The application of 
these results to the early universe requires finite-time
initial conditions. Suppose we take the solution $f_{k(-)}$ as our
initial condition at time $t_0$. When the gravitational background is
curved, the set of adiabatic observers $\{ \tilde f_{k}\}$ 
with
\begin{eqnarray}
{\tilde f}_{k}(t)= { {{[2 \Omega_{k}(t)]}^{-1/2}}}
\exp [-i \int^t dt' \Omega_{k}(t')]
\> ,
\end{eqnarray}
is the
closest we can get to the concept of {\it inertial observers}. In
order to obtain a {\it particle interpretation} we write 
\begin{eqnarray}
f_{k}(t) =
\tilde \alpha_k (t) \tilde f_{k}(t) +
\tilde \beta_k (t) \tilde f_{k}^*(t)
\> . 
\end{eqnarray}
The number of particles is
\begin{eqnarray}
\tilde N_k(t) = \langle 0_{(-)} | \tilde a^{\dag}_k(t) \tilde a_k(t)
|  0_{(-)} \rangle =
{|\tilde \beta_k(t)|}^2
\> , 
\end{eqnarray}
which depends on the
value of $\gamma$ implicitly. We plot the results for $\gamma = 1$ 
and different
values of $k$.

%PUT FIGURE 1 HERE:
\begin{figure}
\vspace{.4cm}
\epsfxsize=8.5cm
\epsfysize=5.5cm
\centerline{\epsfbox{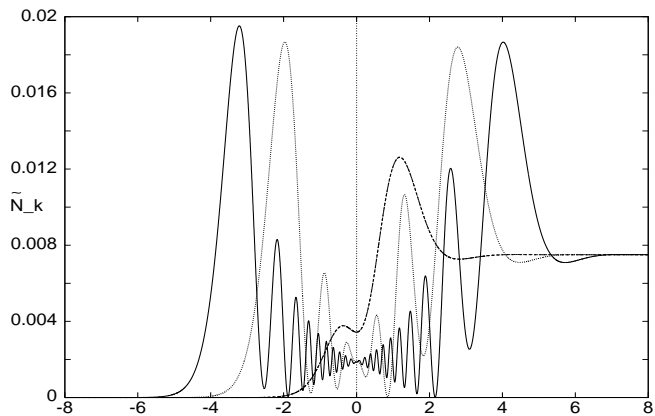}}
\vspace{.3cm}
\caption
{\small{Number of adiabatic particles for different values of
$k$. The dashed, dotted, and solid lines are respectively
$k=0$, $k=5$, and $k=10$.}}
\label{fig:nk}
\end{figure}

\section{Energy-Momentum Tensor}
\label{energymomentum}

The most important physical quantity to compute is the expectation
value of the energy-momentum tensor
in the state defined by the
modes $f_k$. We consider the case $\xi = 1/6$ and $m
\neq 0$. It can be shown that there is a particular choice of
adiabatic observers (those with zeroth adiabatic frequency) for which
the energy
 takes the very simple form
\begin{eqnarray}
{\langle \epsilon \rangle}_{\xi=1/6} 
&=&{1 \over  {(2 \pi^2 a^3)}}
\sum_{k=0}^{+\infty}
{(k+1)}^2
{{\omega_k} \over 2}~   (1+2N_k)~.
\end{eqnarray}
The pressure can be obtained from the conservation equation
\begin{eqnarray}
\dot {\langle \epsilon \rangle}_{\xi=1/6} + 3
 {{\dot a(t)} \over {a(t)}} 
\left[
{\langle \epsilon \rangle}_{\xi=1/6} +
{\langle p \rangle}_{\xi=1/6} 
\right]=0
\> .
\end{eqnarray}
The previous equations can be generalized 
for $\xi \neq 1/6$.

The energy-momentum tensor needs to be regularized and renormalized 
before obtaining its finite physical value.
 We regularize
the energy-momentum tensor by adiabatic methods, since we know that
the fourth order adiabatic energy and pressure have the same
divergences as $\langle \epsilon \rangle$ and $\langle p \rangle$,
respectively~\cite{Bunch}.
We define~\cite{Parker} 
\begin{eqnarray} 
{\langle \epsilon \rangle}_R
=
{\langle \epsilon \rangle}_B
-
{\langle \epsilon \rangle}_A
, \; \; 
{\langle p \rangle}_R
=
{\langle p \rangle}_B
-
{\langle p \rangle}_A
\> . \; \; 
\end{eqnarray}

\section{Conclusions}
\label{sec:conclu}
The study of quantum fields in a curved background requires choosing
appropriate initial conditions, as these
determine the initial state of the quantum system. 
There exists an appropriate adiabatic basis, in which one finds very simple
expressions for the energy and pressure of the quantum
field, which have a classical interpretation: the total energy
of the system is the sum over modes of the number of particles of
momentum $k$ times the frequency of that mode. The pressure
is obtained from the conservation equation.
These results can be generalized for  other values of $\xi$, and
no essential changes in the formalism here developed are needed.

%--------------------------------------------------------------------
\section*{Acknowledgments}
%--------------------------------------------------------------------

This research was supported by the U.S. Department of Energy,
by an NSF Grant and by
Los Alamos National Laboratory. I want to particularly thank
Salman Habib and Emil Mottola for their kind support and wonderful
discussions on the subject.

%--------------------------------------------------------------------
\section*{References}
%--------------------------------------------------------------------

%--------------------------------------------------------------------

\end{document}